\documentclass[12pt]{article}
\usepackage{geometry}                
\usepackage{graphicx}
\usepackage{bm}
\usepackage{amssymb,amsmath,amsthm}
\usepackage{mathrsfs} 
\usepackage{bm}
\usepackage{graphicx,epsfig}
\usepackage[dvips]{color}
\usepackage{amssymb,amsmath,amsthm,amssymb}

\def\J{{\mathscr{J}}}
\def\sech{{\mathrm{sech}}}

\def\TODAY{2 October 2009}

\title{\bf Reformulating the Schr\"odinger equation as a Shabat--Zakharov system}
\author{{\Large Petarpa Boonserm}\\[5pt]
Department of Mathematics, Faculty of Science\\
Chulalongkorn University, Phayathai Rd., Pathumwan\\
Bangkok 10330, Thailand\\[5pt]
{\sf \small petarpa.boonserm@gmail.com} \\[7pt]
{\Large Matt Visser}\\[5pt]
School of Mathematics, Statistics, and Operations Research\\
Victoria University of Wellington, New Zealand\\[5pt]
{\sf \small matt.visser@msor.vuw.ac.nz}  }
\date{\TODAY;  \LaTeX-ed  \today}                                           

\begin{document}
\maketitle
\def\d{{\mathrm{d}}}
\newcommand{\scri}{\mathscr{I}}
\newcommand{\sun}{\ensuremath{\odot}}
\numberwithin{equation}{section} 
\def\J{{\mathscr{J}}}
\def\sech{{\mathrm{sech}}}
\newtheorem{theorem}{Theorem}
\def\Re{{\mathrm{Re}}}
\def\Im{{\mathrm{Im}}}
\begin{abstract}

We reformulate the second-order Schr\"odinger equation as a set of two coupled first order differential equations, a so-called ``Shabat--Zakharov system'', (sometimes called a ``Zakharov--Shabat'' system).  There is considerable flexibility in this approach, and  we emphasise the utility of introducing an  ``auxiliary condition'' or  ``gauge condition'' that is used to cut down the degrees of freedom. Using this formalism, we derive the explicit (but formal) general solution to the Schr\"odinger equation. The general solution depends on three arbitrarily chosen functions,  and a path-ordered exponential matrix.  If one considers path ordering to be an ``elementary'' process,  then this represents complete quadrature, albeit formal, of the second-order linear ODE. 

\enlargethispage{10pt}

\vskip 10 pt


\end{abstract}

\maketitle
\newtheorem{lemma}{Lemma}
\newtheorem{corollary}{Corollary}
\newtheorem{definition}{Definition}
\def\d{{\mathrm{d}}}
\def\implies{\Rightarrow}
\def\arctanh{{\mathrm{arctanh}}}
\def\SIM{\triangleq}
\clearpage
\tableofcontents
\clearpage

\section{Introduction}

The Schr\"odinger equation has now been part of mathematical physics for almost 85 years~\cite{Schrodinger}. Over the years it has been incorporated into many textbook discussions (for a necessarily selective subset see~\cite{Landau, Baym, Gasiorowicz, Capri, Stehle, Schiff, Cohen, Galindo, Park, Fromhold, Scharff, Messiah, Merzbacher, Singh, Mathews}),  it has been the subject of technical research monographs (for a necessarily selective subset see~\cite{Newton0, Newton1, Chadan, Eckhaus, Froman0, Froman1, Froman2, Peierls}), and considerable work has gone into exploring the mathematical foundations of the subject --- see for example~\cite{Langer, Berry0, Berry1, Arnold, Khorasani}. Despite its long and venerable history, foundational questions related to the Schr\"odinger equation still periodically lead to new results~\cite{Bounds0, Bounds-beta, Bounds-greybody, Bounds-mg, Bounds-analytic, Bounds-thesis}.  

In the current article we shall develop a very flexible formalism for reducing the second-order Schr\"odinger equation to a system of two first-order differential equations, a Shabat--Zakharov system.  This generalizes several of our earlier results, in particular those reported by one of the present authors in~\cite{Bounds0}, and provides a formalism for \emph{formally} solving the Schr\"odinger equation in terms of a $2\times2$ ``path-ordered exponential''  matrix. In addition to these formal developments, the technique is notable for the fact that in appropriate circumstances it permits one to derive useful bounds on the behaviour of the exact wave-function.

\section{Basic idea}

Consider the one-dimensional time-independent Schr\"odinger equation~\cite{Schrodinger, Landau, Baym, Gasiorowicz, Capri, Stehle, Schiff, Cohen, Galindo, Park, Fromhold, Scharff, Messiah, Merzbacher, Singh, Mathews}:
\begin{equation}
\label{E:SDE}
-{\hbar^2\over2m} {\d^2\over \d x^2} \psi(x) + V(x) \; \psi(x) = E \; \psi(x).
\end{equation}
Introduce the notation
\begin{equation}
\label{E:k(x)}
k(x)^2 = {2m[E-V(x)]\over\hbar^2}.
\end{equation}
So we are really just trying to solve
\begin{equation}
 {\d^2\over \d x^2} \psi(x) + k(x)^2 \; \psi(x) =0 ,
\end{equation}
or equivalently in the time domain
\begin{equation}
 {\d^2\over \d t^2} \psi(t) + \omega(t)^2 \; \psi(t) =0 .
\end{equation}
Motivated by the JWKB approximation,
\begin{equation}
\label{E:JWKB}
\psi \approx A \; {\exp[i\int k(x)]\over \sqrt{k(x)}} 
+ B \; {\exp[-i\int k(x)]\over \sqrt{k(x)}},
\end{equation}
the key idea is to re-write the second-order Schr\"odinger
equation as a set of two coupled first-order linear differential equations for the coefficients appearing in this linear combination.

Systems of differential equations of this type are often referred to as 
Shabat--Zakharov systems~\cite{Eckhaus, Bounds0, Bounds-thesis}, or sometimes Zakharov--Shabat systems. 
A similar representation of the Schr\"odinger
equation is briefly discussed by Peierls~\cite{Peierls},  and related
representations are well-known, often being used without giving
an explicit reference (see for example~\cite{Bordag}).  However an
exhaustive search has not uncovered prior use of the particular
representation presented here. (Apart, of course, from related precursor work by one of the current authors in~\cite{Bounds0}, and our own more recent related work in~\cite{Bounds-beta, Bounds-greybody, Bounds-mg, Bounds-analytic, Bounds-thesis}.)
Nor, outside of our own work, has any attempt been made to use this Shabat--Zakharov representation to place rigorous \emph{bounds} on the behaviour of 
one-dimensional scattering.

We will start by introducing two arbitrary auxiliary functions $\varphi(x)$ and $\Delta(x)$,
which may at this stage be either real or complex, although we do demand that
$\varphi'(x)\neq 0$, and then define:
\begin{equation}
\label{E:representation}
\psi(x) = 
a(x) \;{\exp(+i \varphi+i\Delta)\over\sqrt{\varphi'}} + 
b(x) \;{\exp(-i \varphi-i\Delta)\over\sqrt{\varphi'}}.
\end{equation}
This representation effectively seeks to use quantities somewhat resembling the ``phase integral'' wavefunctions as a basis for the true wavefunction~\cite{Froman0}. We will ultimately want to interpret $a(x)$ and $b(x)$ as  ``position-dependent JWKB-like coefficients''; in a scattering problem they can be thought of as
``position-dependent Bogoliubov coefficients''. The representation given above is of course extremely highly redundant:  \emph{one} complex number $\psi(x)$ has been traded for \emph{two} complex numbers $a(x)$ and $b(x)$, \emph{plus} two essentially arbitrary auxiliary functions $\varphi(x)$ and $\Delta(x)$.  
To reduce this freedom, or more precisely keep it firmly under control,  we introduce an ``auxiliary condition'' (or ``auxiliary constraint'', or ``gauge condition''):
\begin{equation}
\label{E:gauge-shabat}
{\d\over \d x}\left({a \exp(i \Delta) \over\sqrt{\varphi'}}\right) e^{+i \varphi} + 
{\d\over \d x}\left({b \exp(-i\Delta) \over\sqrt{\varphi'}}\right) e^{-i \varphi} 
= \chi(x) \; \psi(x).
\end{equation}
Here $\chi(x)$ is yet a third arbitrary function of position. It is allowed
to be complex, and may be zero. The original analysis, published in~\cite{Bounds0} corresponds to the
special case $\Delta(x)=0$ and $\chi(x)=0$, so that it clear that the current analysis is a significant generalization. Subject to this ``gauge condition'', it is easy to evaluate: 
\begin{equation}
\label{E:gradient}
{\d\psi\over \d x} = i \sqrt{\varphi'} 
\left\{ a(x) \exp(+i \varphi+i\Delta) - b(x) \exp(-i \varphi-i\Delta ) \right\} + \chi\; \psi.
\end{equation}
Repeated differentiation of this equation will soon lead to our desired result.

\section{Probability current}
To give us some insight into the physical meaning of the coefficients $a(x)$ and $b(x)$ it is useful to first calculate the probability current.
As usual we take
\begin{equation}
\label{newdensity}
\mathscr{J} (x,t) = {\hbar \over 2 m  \, i} \, \bigg(\psi^* {\partial \psi \over \partial x} - {\partial \psi^* \over \partial x} \, \psi\bigg)
= {\hbar \over  m} \, \mathrm{Im} \bigg(\psi^* {\partial \psi \over \partial x}\bigg).
\end{equation}
Here $(\hbar / m)$ is just a normalization (that is often set $\rightarrow 1$ for 
convenience). There is nothing really important in this normalization (unless we want to calculate 
experimental numbers), so we might as well set
\begin{equation}
\mathscr{J} (x,t) =  \mathrm{Im} \bigg(\psi^* {\partial \psi \over \partial x}\bigg).
\end{equation}
Using our JWKB-based ansatz in terms of $a(x)$ and $b(x)$ we compute:
\begin{eqnarray}
\mathscr{J} 
&=& \mathrm{Im} \, \bigg\{ \psi^*\left[i \sqrt{\varphi'}\{a(x) \exp (+i \varphi+i\Delta) - b(x) \exp (-i \varphi-i\Delta)\} + \chi \, \psi \right] \bigg\} ,\qquad
\end{eqnarray}
whence
\begin{eqnarray}
\mathscr{J}  &=& \mathrm{Re} \, \Bigg\{\sqrt{{\varphi' \over \varphi'^*}} \; [a(x) \exp (+i \varphi+i\Delta) - b(x) \exp (-i \varphi-i\Delta)]
\nonumber
\\
&&
\times [a(x)^* \exp (-i \varphi^*-i\Delta^*) + b(x)^* \exp (+ i \varphi^*+i\Delta^*)] \Bigg\} 
+ 
\mathrm{Im} \{\chi\} \, \psi^* \psi.\qquad
\end{eqnarray}
This implies
\begin{eqnarray}
\mathscr{J}  &=& \mathrm{Re} \, \Bigg\{\sqrt{{\varphi' \over \varphi'^*}}\Bigg\} \; \left[|a|^2 \mathrm{Re}\{e^{+i(\varphi+\Delta-\varphi^*-\Delta^*)}\}
 - |b|^2 \mathrm{Re}\{e^{-i(\varphi+\Delta-\varphi^*-\Delta^*)}\}\right] 
\nonumber\\ &&
+ \mathrm{Im} \, \Bigg\{\sqrt{\varphi' \over \varphi'^*} \Bigg\} \; \mathrm{Im}\left\{ab^* \, e^{i(\varphi +\Delta+ \varphi^*+\Delta^*)}\right\} + \mathrm{Im} \{ \chi\} \, \psi^* \psi  ,
\end{eqnarray}
which we can finally recast as
\begin{eqnarray}
\mathscr{J}  
&=&   {\mathrm{Re}\{\varphi'\} \over |\varphi'|} \left[|a|^2\mathrm{Re} \{e^{+2 \mathrm{Im}(\varphi+\Delta)}\} - |b|^2 \mathrm{Re}\{e^{-2 \mathrm{Im}(\varphi+\Delta)}\}\right] 
\nonumber\\ &&
+ {\mathrm{Im}\{\varphi'\} \over |\varphi'|} \, \mathrm{Im} \{a b ^* \, e^{2 i \, \mathrm{Re}(\varphi+\Delta)}\} + \mathrm{Im} \{\chi\} \, \psi^* \psi.
\end{eqnarray}
Recall that at this stage $\varphi(x)$, $\Delta(x)$, and  $\chi(x)$ are completely arbitrary possibly complex functions subject only to the constraint $\varphi' \neq 0$. 

If we now temporarily demand that  $\varphi(x)$, $\Delta(x)$, and  $\chi(x)$ are real we see that
\begin{equation}
\mathscr{J}  \rightarrow |a|^2 - |b|^2,
\end{equation}
an observation that strongly suggests that at least in those circumstances the quantities $a(x)$ and $b(x)$ might usefully be thought of as ``position-dependent Bogoliubov coefficients''. 

\section{Schr\"odinger equation  as a first order system}
%
We shall now re-write the Schr\"odinger equation in terms of two coupled
first-order differential equations for these position-dependent
JWKB/Bogoliubov coefficients  $a(x)$ and $b(x)$. To do this we evaluate the quantity  $\d^2\psi/ \d x^2$ in two different ways, making
repeated use of the gauge condition. From
\begin{eqnarray}
\label{E:double-gradient}
{\d^2\psi\over \d x^2} 
&=& 
{\d\over \d x} 
\left( i{\varphi'\over \sqrt{\varphi'}} 
\left\{ a e^{+i \varphi+i\Delta} - b e^{-i \varphi-i\Delta} \right\}  + \chi\;\psi
\right),
\end{eqnarray}
we first see
\begin{eqnarray}
{\d^2\psi\over \d x^2}  &=& {(i\varphi')^2\over\sqrt{\varphi'}} 
\left\{ a e^{+i \varphi+i\Delta} + b e^{-i \varphi-i\Delta} \right\}
\nonumber\\
&&+ i \varphi'
\left\{ 
{\d\over \d x}\left({a\,e^{i\Delta}\over\sqrt{\varphi'}}\right) e^{+i \varphi} - 
{\d\over \d x}\left({b\, e^{-i\Delta}\over\sqrt{\varphi'}}\right) e^{-i \varphi} 
\right\} 
\nonumber\\
&&
+i{\varphi''\over\sqrt{\varphi'}} 
\left\{ a e^{+i \varphi+i\Delta} - b e^{-i \varphi-i\Delta} \right\}
+ \chi' \; \psi + \chi\; \psi' ,
\end{eqnarray}
so that
\begin{eqnarray}
{\d^2\psi\over \d x^2} &=& -{\varphi'^2\over\sqrt{\varphi'}}  
\left\{ a e^{+i \varphi+i\Delta} + b e^{-i \varphi-i\Delta} \right\}
\nonumber\\
&&+i\varphi' \left\{ 2
{\d\over \d x}\left({a\,e^{i\Delta}\over\sqrt{\varphi'}}\right) e^{+i \varphi} - \chi \psi 
\right\}
\nonumber\\
&&
+i{\varphi''\over\sqrt{\varphi'}} 
\left\{ a e^{+i \varphi+i\Delta} - b e^{-i \varphi-i\Delta} \right\} + \chi' \; \psi + \chi\; \psi'.
\end{eqnarray}
But then
\begin{eqnarray}
\label{E:double-gradient-b}
{\d^2\psi\over \d x^2} 
&=& -\varphi'^2 \; \psi
+{2 i\varphi'\over\sqrt{\varphi'}}  {\d a\over \d x} e^{+i \varphi+i\Delta} 
-2\sqrt{\varphi'} \Delta' e^{i\varphi+i\Delta} a 
\nonumber\\
&& - i{\varphi''\over\sqrt{\varphi'}} b e^{-i \varphi-i\Delta} 
-i \varphi' \chi \psi
+ \chi' \psi 
\nonumber\\
&& + \chi\; \left[ i \sqrt{\varphi'} 
\left\{ a(x) e^{+i \varphi+i\Delta} - b(x) e^{-i \varphi-i\Delta} \right\} + \chi\; \psi \right].
\end{eqnarray}
So finally
\begin{eqnarray}
\label{E:double-gradient-c}
{\d^2\psi\over \d x^2} 
&=& \left[\chi^2+\chi'-(\varphi')^2\right] \psi
+{2 i\varphi'\over\sqrt{\varphi'}}  {\d a\over \d x} e^{+i \varphi+i\Delta} 
\nonumber\\
&&-2\sqrt{\varphi'} \Delta' e^{i\varphi+i\Delta} a 
-i{[\varphi''+2\chi\varphi']\over\sqrt{\varphi'}} b e^{-i \varphi-i\Delta}.
\nonumber\\
&&
\end{eqnarray}
Now use the gauge condition to eliminate $\d a/\d x$ in favour of $\d
b/\d x$.  This permits us to write the quantity $\d^2\psi/ \d x^2$ in
either of the two equivalent forms
\begin{eqnarray}
\label{E:double-gradient2}
{\d^2\psi\over \d x^2} 
&=& 
\left[\chi^2+\chi'-(\varphi')^2\right] \psi 
- 2 i \varphi' {\d b\over \d x} {e^{-i\varphi-i\Delta}\over\sqrt{\varphi'}} 
\nonumber \\
&&
-2\sqrt{\varphi'} \Delta' e^{-i\varphi-i\Delta} b 
+ i \left[ \varphi''+2\chi \varphi' \right] a {e^{+i\varphi+i\Delta}\over\sqrt{\varphi'}},
\end{eqnarray}
and/or
\begin{eqnarray}
\label{E:double-gradient2b}
{\d^2\psi\over \d x^2} 
&=& 
\left[\chi^2+\chi'-(\varphi')^2\right] \psi 
+ 2 i \varphi' {\d a\over \d x} {e^{+i\varphi+i\Delta}\over\sqrt{\varphi'}}
\nonumber
\\
&&
-2\sqrt{\varphi'} \Delta' e^{i\varphi+i\Delta} a 
- i \left[ \varphi''+2\chi \varphi' \right] b {e^{-i\varphi-i\Delta}\over\sqrt{\varphi'}}. 
\end{eqnarray}
We now insert these formulae into the Schr\"odinger equation written in the
form
\begin{equation}
{\d^2\psi\over \d x^2} + k(x)^2 \; \psi  = 0,
\end{equation}
to deduce the first-order system:
\begin{eqnarray}
\label{E:system-a}
\!\!\!
{\d a\over \d x} &=& + {1\over 2\varphi'} \Bigg\{ 
i\left[k^2(x)+\chi^2+\chi'-(\varphi')^2-2\varphi'\Delta'\right]  \; a
\nonumber \\
&&
+\left([\varphi''+2\chi\varphi']+ i\left[k^2(x)+\chi^2+\chi'-(\varphi')^2\right]\right) \; e^{-2i\varphi-2i\Delta} \; b
\Bigg\},\;
\\
\label{E:system-b}
\!\!\!
{\d b\over \d x} &=&  +{1\over 2\varphi'} \Bigg\{ 
\left( [\varphi''+2\chi\varphi'] - i\left[k^2(x)+\chi^2+\chi'-(\varphi')^2\right]  \right)\; 
e^{+2i\varphi+2i\Delta}  \; a
\nonumber \\
&&
- i\left[k^2(x)+\chi^2+\chi'-(\varphi')^2-2\varphi'\Delta'\right]  \; b
\Bigg\}.
\end{eqnarray}
It is easy to verify that this first-order system is compatible with
the ``gauge condition'' (\ref{E:gauge-shabat}), and that by iterating the
system twice (subject to this gauge condition) one recovers exactly
the original Schr\"odinger equation. These equations hold for arbitrary
$\varphi(x)$, $\Delta(x)$, and $\chi(x)$, real or complex (subject only to $\varphi'\neq0$ to avoid divide by zero issues). 

This system can compactly be written in $2\times2$ matrix
form, and once this has been done these equations exhibit a deep connection with the transfer matrix
formalism~\cite{Transfer}.  Let us define quantities
$\rho_1(x)$ and $\rho_2(x)$, not necessarily real, as
\begin{equation}
\rho_1 = \varphi''+2\chi \varphi';
\qquad
\rho_2 = k^2(x)+\chi^2+\chi'-(\varphi')^2.
\end{equation}
We then re-write the Shabat--Zakharov system in $2\times2$ matrix form as
\begin{equation}
{\d\over \d x} \left[\begin{matrix} a \\ b\end{matrix}\right] = 
{1\over2\varphi'} 
\left[
\begin{matrix}
i [\rho_2-2\varphi'\Delta']   & 
\left\{ \rho_1+i\rho_2\right\} \exp(-2i\varphi-2i\Delta) \\
\left\{ \rho_1-i\rho_2\right\}  \exp(+2i\varphi+2i\Delta)  & 
-i[\rho_2-2\varphi'\Delta']
\end{matrix}
\right]
\left[ \begin{matrix} a \cr b\end{matrix}\right].
\end{equation}
This has the formal solution
\begin{equation}
\left[\begin{matrix} a(x) \cr b(x)\end{matrix}\right] = 
E(x,x_0)  \;\left[\begin{matrix} a(x_0) \cr b(x_0)\end{matrix}\right],
\end{equation}
in terms of a generalized position-dependent ``transfer
matrix''~\cite{Transfer}
\begin{eqnarray}
&&
E(x,x_0) = 
\nonumber
\\
&&
\qquad {\cal P} \exp\left( 
\int_{x_0}^{x}
{1\over2\varphi'} 
\left[
\begin{matrix}
i [\rho_2-2\varphi'\Delta']   & 
\left\{ \rho_1+i\rho_2 \right\} \; e^{-2i\varphi-2i\Delta} \cr 
\left\{ \rho_1-i\rho_2 \right\}  \; e^{+2i\varphi+2i\Delta}  & 
-i[\rho_2-2\varphi'\Delta']
\end{matrix}
\right] 
 \d x \right),
\nonumber
\\
&&
\end{eqnarray}
where the symbol ${\cal P}$ denotes ``path ordering''. 

Equivalently, if we were to be working in the time domain we would have 
\begin{eqnarray}
&&
E(t,t_0) = 
\nonumber
\\
&&
\qquad {\cal T} \exp\left( 
\int_{t_0}^{t}
{1\over2\dot\varphi} 
\left[
\begin{matrix}
i [\rho_2-2\dot\varphi\,\dot\Delta]   & 
\left\{ \rho_1+i\rho_2 \right\} \; e^{-2i\varphi-2i\Delta} \cr 
\left\{ \rho_1-i\rho_2 \right\}  \; e^{+2i\varphi+2i\Delta}  & 
-i[\rho_2-2\dot\varphi\,\dot\Delta]
\end{matrix}
\right] 
 \d t \right),
\nonumber
\\
&&
\end{eqnarray}
where ${\cal T}$ would now be the well-known ``time ordering'' operator (more usually encountered in a quantum field theory setting) and we would now define
\begin{equation}
\rho_1 = \ddot\varphi+2\chi \dot\varphi;
\qquad
\rho_2 = \omega^2(t)+\chi^2-\dot\chi-(\dot\varphi)^2,
\end{equation}
with $\varphi(t)$, $\Delta(t)$, and $\chi(t)$ now being arbitrary functions of $t$ rather than $x$, and $k(x) \rightarrow \omega(t)$.

Returning to position space, we can now write the (exact) wave function in
inner product form
\begin{equation}
\psi(x) =   
{1\over\sqrt{\varphi'}}
\left[\begin{matrix}
{\exp(+i \varphi+i\Delta)}; & 
{\exp(-i \varphi-i\Delta)}
\end{matrix} \right] \;
\left[\begin{matrix} a(x) \cr b(x)\end{matrix}\right],
\end{equation}
to yield a \emph{formal but completely general} solution for the Schr\"odinger
equation
\begin{equation}
\psi(x) = {1\over\sqrt{\varphi'}} 
\left[\begin{matrix}
{\exp(+i \varphi+i\Delta)}; & 
{\exp(-i \varphi-i\Delta)}
\end{matrix} \right] 
E(x,x_0)  
\left[\begin{matrix} a(x_0) \cr b(x_0)\end{matrix}\right].
\end{equation}
Explicitly
\begin{eqnarray}
&&
\psi(x) = 
{1\over\sqrt{\varphi'}}
\left[\begin{matrix}
{\exp(+i \varphi+i\Delta)}; & 
{\exp(-i \varphi+i\Delta)}
\end{matrix} \right]
\\
&&
\quad
\times
{\cal P} \exp\left( 
\int_{x_0}^{x}
{1\over2\varphi'} 
\left[
\begin{matrix}
i [\rho_2-2\varphi'\Delta']   & 
\left\{ \rho_1+i\rho_2 \right\} \; e^{-2i\varphi-2i\Delta} \cr 
\left\{ \rho_1-i\rho_2 \right\}  \; e^{+2i\varphi+2i\Delta}  & 
-i[\rho_2-2\varphi'\Delta']
\end{matrix}
\right] 
\d \bar x \right)
\left[\begin{matrix} a(x_0) \cr b(x_0)\end{matrix}\right].
\nonumber
\end{eqnarray}
This is the explicit general solution to the Schr\"odinger equation. It
depends on the three arbitrarily chosen and possibly complex functions $\varphi(x)$, $\Delta(x)$, and
$\chi(x)$,  and a path-ordered exponential matrix.  If you consider path
ordering to be an ``elementary'' process, then this is a complete quadrature, albeit formal, of the Schr\"odinger equation, and implicitly, of the general
second-order linear ODE.

\section{Special cases}

We can now use the freedom in choosing $\varphi(x)$, $\Delta(x)$, and $\chi(x)$ to explore some special cases where the Shabat--Zakharov system simplifies.

\subsection{Case: $\Delta' = \rho_2/(2\varphi')$}

No one can prevent us from choosing
\begin{equation}
\Delta' = {\rho_2\over2\varphi'},
\end{equation}
that is
\begin{equation}
\Delta' = {k^2(x)+\chi^2+\chi'-(\varphi')^2\over2\varphi'},
\end{equation}
which implies
\begin{equation}
\Delta = \int {k^2(x)+\chi^2+\chi'-(\varphi')^2\over2\varphi'} \; \d x.
\end{equation}
Doing this greatly simplifies the Shabat--Zakharov system  since now 
\begin{eqnarray}
\label{E:system-a1}
{\d a\over \d x} &=& +
{1\over 2\varphi'} 
\Bigg\{ 
\left([\varphi''+2\chi\varphi']+ i\left[k^2(x)+\chi^2+\chi'-(\varphi')^2\right]\right) \; e^{-2i\varphi-2i\Delta} 
\Bigg\} \; b,
\nonumber
\\
&&
\\
\label{E:system-b1}
{\d b\over \d x} &=& +
{1\over 2\varphi'} 
\Bigg\{ 
\left( [\varphi''+2\chi\varphi'] - i\left[k^2(x)+\chi^2+\chi'-(\varphi')^2\right]  \right)\; 
e^{+2i\varphi+2i\Delta}  
\Bigg\}\; a.
\nonumber
\\
&&
\end{eqnarray}
That is
\begin{equation}
{\d\over \d x} \left[\begin{matrix} a \\ b\end{matrix}\right] = 
{1\over2\varphi'} 
\left[
\begin{matrix}
0 & 
\left\{ \rho_1+i\rho_2\right\} \exp(-2i\varphi-2i\Delta) \\
\left\{ \rho_1-i\rho_2\right\}  \exp(+2i\varphi+2i\Delta)  & 
0
\end{matrix}
\right]
\left[ \begin{matrix} a \cr b\end{matrix}\right].
\end{equation}
In this situation one has eliminated the diagonal part of the metric --- the cost unfortunately is that the off-diagonal components will now have a rapidly oscillating phase. (A somewhat related pair of equations can be found in reference~\cite{Berry1}.)

\subsection{Case: $\Delta=-\varphi$}

No one can prevent us from choosing
\begin{equation}
\Delta(x) = -\varphi(x),
\end{equation}
in which case
\begin{eqnarray}
\label{E:system-a2}
{\d a\over \d x} &=& +
{1\over 2\varphi'} 
\Bigg\{ 
i\left[k^2(x)+\chi^2+\chi'+(\varphi')^2\right]  \; a
\nonumber\\
&&
+
\left([\varphi''+2\chi\varphi']+ i\left[k^2(x)+\chi^2+\chi'-(\varphi')^2\right]\right)  \; b
\Bigg\},
\\
\label{E:system-b2}
{\d b\over \d x} &=& +
{1\over 2\varphi'} 
\Bigg\{ 
\left( [\varphi''+2\chi\varphi'] - i\left[k^2(x)+\chi^2+\chi'-(\varphi')^2\right]  \right)\;   \; a
\nonumber\\
&&
- i\left[k^2(x)+\chi^2+\chi'+(\varphi')^2\right]  \; b
\Bigg\}.
\end{eqnarray}
The complicated phase structure has gone away, and we now have 
\begin{equation}
{\d\over \d x} \left[\begin{matrix} a \\ b\end{matrix}\right] = 
{1\over2\varphi'} 
\left[
\begin{matrix}
i [\rho_2+2(\varphi')^2]   & 
\left\{ \rho_1+i\rho_2\right\} \\
\left\{ \rho_1-i\rho_2\right\} & 
-i[\rho_2+2(\varphi')^2]
\end{matrix}
\right]
\left[ \begin{matrix} a \cr b\end{matrix}\right].
\end{equation}
The phases of the matrix entries are now slowly varying --- the price to pay is that there is a full complement of nonzero matrix entries to deal with.

\subsection{Case: $\Delta=0$}

We include this case mainly for historical reasons, as it is an otherwise unpublished result that was the first significant generalization we obtained of the original result published in~\cite{Bounds0}. 
The Shabat--Zakharov system in this case simplifies to
\begin{equation}
{\d\over \d x} \left[\begin{matrix} a \\ b\end{matrix}\right] = 
{1\over2\varphi'} 
\left[
\begin{matrix}
i \rho_2   & 
\left\{ \rho_1+i\rho_2\right\} \exp(-2i\varphi) \\
\left\{ \rho_1-i\rho_2\right\}  \exp(+2i\varphi)  & 
-i[\rho_2-2\varphi]
\end{matrix}
\right]
\left[ \begin{matrix} a \cr b\end{matrix}\right].
\end{equation}
Since on still has the freedom to choose both $\varphi$ and $\chi$ this is definitely more general than~\cite{Bounds0}, that article corresponding to the specialization $\chi\to 0$.


\section[Application: \\ Bounding the coefficients $a(x)$ and $b(x)$]{Application: \\ Bounding the coefficients $a(x)$ and $b(x)$}

One of the particuarly  interesting applications we have found for the Shabat--Zakharov system derived above is that it is possible to use it to place quite general and rigourous bounds on the coefficients evolution $a(x)$ and $b(x)$. 
From the general Shabat--Zakharov system
\begin{eqnarray}
\label{E:system-a3}
{\d a\over \d x} &=& +
{1\over 2\varphi'} 
\Bigg\{ 
i\left[k^2(x)+\chi^2+\chi'-(\varphi')^2-2\varphi'\Delta'\right]  \; a
\nonumber\\
&&
\qquad
+
\left([\varphi''+2\chi\varphi']+ i\left[k^2(x)+\chi^2+\chi'-(\varphi')^2\right]\right) \; e^{-2i\varphi-2i\Delta} \; b
\Bigg\},
\nonumber
\\
&&
\\
\label{E:system-b3}
{\d b\over \d x} &=& +
{1\over 2\varphi'} 
\Bigg\{ 
\left( [\varphi''+2\chi\varphi'] - i\left[k^2(x)+\chi^2+\chi'-(\varphi')^2\right]  \right)\; 
e^{+2i\varphi+2i\Delta}  \; a
\nonumber\\
&&
\qquad
- i\left[k^2(x)+\chi^2+\chi'-(\varphi')^2-2\varphi'\Delta'\right]  \; b
\Bigg\},
\end{eqnarray}
we see
\begin{eqnarray}
a^* {\d a\over \d x} &=& +
{1\over 2\varphi'} 
\Bigg\{ 
i\left[k^2(x)+\chi^2+\chi'-(\varphi')^2-2\varphi'\Delta'\right]  \; a^* a
\nonumber\\
&&
\quad
+
\left([\varphi''+2\chi\varphi']+ i\left[k^2(x)+\chi^2+\chi'-(\varphi')^2\right]\right) \; e^{-2i\varphi-2i\Delta} \; a^* \; b
\Bigg\}.
\nonumber
\\
&&
\end{eqnarray}
Therefore
\begin{eqnarray}
a^* {\d a\over \d x} + a {\d a^*\over \d x} &=&
\Im\left\{ { k^2(x)+\chi^2+\chi'-(\varphi')^2-2\varphi'\Delta' \over \varphi' } \right\}   a^* a
\nonumber
\\
&+& \Re\left\{ {\left([\varphi''+2\chi\varphi']+ i\left[k^2(x)+\chi^2+\chi'-(\varphi')^2\right]\right) \; e^{-2i\varphi-2i\Delta} \; a^* \; b\over\varphi'} \right\} .
\nonumber
\\
&&
\end{eqnarray}
But $\Re(A)\leq |A|$, whence
\begin{eqnarray}
 {\d |a|^2\over \d x}  &\leq&
\Im\left\{ { k^2(x)+\chi^2+\chi'-(\varphi')^2-2\varphi'\Delta' \over \varphi' } \right\}   |a|^2 
\nonumber
\\
&+& \left| {\left([\varphi''+2\chi\varphi']+ i\left[k^2(x)+\chi^2+\chi'-(\varphi')^2\right]\right) \; e^{-2i\varphi-2i\Delta} \; a^* \; b\over\varphi'} \right| .
\nonumber
\\
&&
\end{eqnarray}
Therefore
\begin{eqnarray}
 {\d |a|\over \d x}  &\leq&
\Im\left\{ { k^2(x)+\chi^2+\chi'-(\varphi')^2-2\varphi'\Delta' \over 2 \varphi' } \right\}  \;  |a|
\nonumber
\\
&+& \left| {[\varphi''+2\chi\varphi']+ i\left[k^2(x)+\chi^2+\chi'-(\varphi')^2\right] \over 2 \varphi'} \right| \; e^{2\Im(\varphi+\Delta)} \; |b| .
\nonumber
\\
&&
\end{eqnarray}
While up to this stage  $\varphi$, $\Delta$, and $\chi$ have been allowed to be complex, we have found that for current purposes (establishing the bounds) it proves impractical to retain this level of generality, and to make any progress we must restrict attention to real $\varphi$, $\Delta$, and $\chi$.   
The inequality now reduces to
\begin{equation}
{\d |a|\over \d x} \leq  \left| {\; [\varphi''+2\chi\varphi']+ i\left[k^2(x)+\chi^2+\chi'-(\varphi')^2\right]  \over2\varphi'} \right| \; |b|,
\end{equation}
and so
\begin{equation}
{\d |a|\over \d x} \leq  {\sqrt{ [\varphi''+2\chi\varphi']^2 + \left[k^2(x)+\chi^2+\chi'-(\varphi')^2\right]^2} \over2|\varphi'|}  \; |b|.
\end{equation}
We note that $\Delta$ has now completely disappeared from the inequality. 
Under the current assumptions it is easy to check that
\begin{equation}
\mathscr{J} = \Im(\psi^* \psi') = |a|^2- |b|^2,
\end{equation}
so current conservation implies
\begin{equation}
\label{the-current-conservation}
|a|^2- |b|^2 = 1.
\end{equation}
(Ultimately, it is this equation that allows us to interpret $a(x)$ and $b(x)$ as ``position-dependent Bogoliubov coefficients''.)
In view of this relation between $a(x)$ and $b(x)$  we have $|b| = \sqrt{|a|^2-1}$, so that we can deduce
\begin{equation}
{\d |a|\over \d x} \leq  {\sqrt{ [\varphi''+2\chi\varphi']^2 + \left[k^2(x)+\chi^2+\chi'-(\varphi')^2\right]^2} \over2|\varphi'|} \; \sqrt{|a|^2-1}.
\end{equation}
But this inequality can now be integrated. For convenience let us define
\begin{equation}
\label{E:theta}
\vartheta =  {\sqrt{ [\varphi''+2\chi\varphi']^2 + \left[k^2(x)+\chi^2+\chi'-(\varphi')^2\right]^2} \over2|\varphi'|}.
\end{equation}
Then 
\begin{equation}
{\d |a|\over \d x} \leq \vartheta \; \sqrt{|a|^2-1}.
\end{equation}
But now
\begin{equation}
\int {1\over  \sqrt{|a|^2-1} } \; {\d |a|\over \d x} \; \d x \leq \int \vartheta \; \d x,
\end{equation}
so that
\begin{equation}
\left\{ \cosh^{-1} |a| \right\}_{x_i}^{x_f}  \leq \int_{x_i}^{x_f}  \vartheta \; \d x.
\end{equation}
Now apply suitable boundary conditions: as $x_i\to-\infty$ we can choose to set things up so that we have a pure transmitted wave, so $|b(-\infty)|=0$ and $|a(-\infty)|=1$.  On the other hand as  $x_f\to+\infty$ we must then choose to set things up so that $a(x)$ and $b(x)$ tend to $\alpha$ and $\beta$, the Bogoliubov coefficients we are interested in calculating. Thus taking the double limit  $x_i\to-\infty$  and  $x_f\to+\infty$  we see:
\begin{equation}
\cosh^{-1} |\alpha|  \leq \int_{-\infty}^{+\infty}  \vartheta \; \d x.
\end{equation}
That is
\begin{equation}
\label{E:alpha}
|\alpha|  \leq \cosh \left\{ \int_{-\infty}^{+\infty}  \vartheta \; \d x\right\} .
\end{equation}
This is the central result of this article --- it can be modified and rearranged in a number of ways, and related inequalities can be derived under slightly different hypotheses, but all the applications we are interested in will reduce in one way or another to an application of this inequality or one of its close variants.

For notational convenience, we often find it is useful to adopt the shorthand
\begin{equation}
\oint   =  \int_{-\infty}^{+\infty} ,
\end{equation}
since then
\begin{equation}
|\alpha|  \leq \cosh \left\{ \oint  \vartheta \; \d x\right\} .
\end{equation}
From the normalization condition (\ref{the-current-conservation}) we immediately deduce
\begin{equation}
|\beta|  \leq \sinh \left\{ \oint  \vartheta \; \d x\right\} .
\end{equation}
When translated into equivalent statements about transmission and reflection probabilities, we find
\begin{equation}
T  \geq \sech^2 \left\{ \oint  \vartheta \; \d x\right\} ,
\end{equation}
and
\begin{equation}
R  \leq \tanh^2 \left\{ \oint  \vartheta \; \d x\right\} .
\end{equation}
where we reiterate
\begin{equation}
\oint \vartheta \; \d x =  \oint {\sqrt{ [\varphi''+2\chi\varphi']^2 + \left[k^2(x)+\chi^2+\chi'-(\varphi')^2\right]^2} \over2|\varphi'|} \; \d x.
\end{equation}
The equivalent bound in the case $\chi=0$ was previously derived in~\cite{Bounds0}, and via a rather different technique verified in~\cite{Bounds-beta}.  The current bound is definitely stronger than anything reported in \cite{Bounds0, Bounds-beta}, though somewhat surprisingly it can (after some transformations) be shown to be equivalent to the bound derived in~\cite{Bounds-mg} by using a radically different technique involving the Miller--Good transformation. Be that as it may, the underlying Shabat--Zakharov system is ultimately of deeper significance and we continue to investigate the possibility of deriving  improved bounds using the current and related techniques.


\section{Discussion}
There are several ways of recasting the Schr\"odinger equation in a form where it is more amenable to formal analysis. In this article we have rewritten the Schr\"odinger equation in terms of an equivalent system of first-order equations --- a Shabat--Zakharov  system --- and then analytically studied this system. In particular we have used the system to derive a number of rigourous bounds on transmission probabilities (and reflection probabilities and Bogoliubov coefficients) for one-dimensional scattering problems, and compared them with earlier results in~\cite{Bounds0, Bounds-beta, Bounds-mg}. 

Even though the calculations we have presented are sometimes somewhat tedious, we feel that they are more than worth the effort --- since there is a fundamental lesson to be learnt from them. Technically, we demonstrated that the Schr\"odinger equation can be written as a Shabat--Zakharov system, which  can then be re-written in $2\times2$ matrix form. We explicitly derived the general solution  in terms of a position-dependent ``transfer matrix''  involving  the symbol $\mathcal{P}$ which denotes ``path ordering''. This explicit general solution to the Schr\"odinger equation depends on the three arbitrarily chosen functions $\varphi(x)$, $\Delta(x)$,  and $\chi(x)$ and a path-ordered exponential matrix. If one considers path ordering to be an ``elementary'' process, then this is the holy grail of ODE theory (complete quadrature, albeit formal, of the second-order linear ODE).

\section*{Acknowledgments}
This research was supported by the Marsden Fund administered by the Royal Society of New Zealand. PB was additionally supported by a scholarship from the Royal Government of Thailand. 



\begin{thebibliography}{69}

\bibitem{Schrodinger}
Schr\"odinger, Erwin. ``An Undulatory Theory of the Mechanics of Atoms and Molecules". 
Phys. Rev. 28 (1926): 1049--1070. doi:10.1103/PhysRev.28.1049.


\bibitem{Landau}
L.~D.~Landau and E.M.~Lifshitz,
\emph{Quantum Mechanics: Non-relativistic theory}, (Pergamon, New York, 1977).

\bibitem{Baym}
G.~Baym,
\emph{Lectures on Quantum Mechanics}, (Benjamin, New York, 1969).

\bibitem{Gasiorowicz}
S.~Gasiorowicz,
\emph{Quantum Physics}, (Wiley, New York, 1996).

\bibitem{Capri}
A.~Z.~Capri,
\emph{Non-relativistic Quantum Mechanics}, 
(Benjamin-Cummings, Menlo Park, California, 1985). See esp. pp. 95-109.

\bibitem{Stehle}
P.~Stehle,
\emph{Quantum Mechanics}, 
(Holden-Day, San Francisco, 1996). See esp. pp. 57-60.

\bibitem{Schiff}
L.~I.~Schiff,
\emph{Quantum Mechanics}, 
(McGraw-Hill, New York, 1955).

\bibitem{Cohen}
C.~Cohen-Tannoudji,~B.~Dui, and F.~Lalo\"e,
\emph{Quantum Mechanics}, 
(Wiley, New York, 1977).

\bibitem{Galindo}
A.~Galindo and P.~Pascual,
\emph{Quantum Mechanics I}, 
(Springer-Verlag, Berlin, 1990).

\bibitem{Park}
D.~Park,
\emph{Introduction to the Quantum Theory}, 
(McGraw-Hill, New York, 1974).

\bibitem{Fromhold}
A.~T.~Fromhold,
\emph{Quantum mechanics for applied physics and engineering}, 
(Academic, New York, 1981).

\bibitem{Scharff}
M.~Scharff,
\emph{Elementary Quantum Mechanics}, 
(Wiley, London, 1969).

\bibitem{Messiah}
A.~Messiah,
\emph{Quantum Mechanics},
(North-Holland, Amsterdam, 1958).

\bibitem{Merzbacher}
E.~Merzbacher,
\emph{Quantum Mechanics}, (Wiley, New York, 1965).

\bibitem{Singh}
J.~Singh,
\emph{Quantum Mechanics: Fundamentals and applications to technology}, 
(Wiley, New York, 1997).

\bibitem{Mathews}
P.~M.~Mathews and K.~Venkatesan,
\emph{A textbook of Quantum Mechanics}, 
(McGraw-Hill, New York, 1978).




\bibitem{Newton0}
R.~G.~Newton,
\emph{Scattering Theory of Waves and Particles},
(McGraw--Hill, New York, 1965).

\bibitem{Newton1}
R.~G.~Newton,
\emph{Inverse Schrodinger Scattering in Three Dimensions},
(Springer, New York, 1990).

\bibitem{Chadan}
K.~Chadan and P.~C.~Sabatier,
\emph{Inverse problems in quantum scattering theory},
(Springer-Verlag, New York, 1989).

\bibitem{Eckhaus}
W.~Eckhaus and A.~Van~Harten,
\emph{The Inverse Scattering Transformation and the theory of Solitons},
(North-Holland, Amsterdam, 1981).


\bibitem{Froman0}
N.~ Froman and P.~O.~Froman,
\emph{JWKB Approximation: Contributions to the Theory},
(North-Holland, Amsterdam, 1965).

\bibitem{Froman1}
N.~ Froman and P.~O.~Froman,
\emph{Phase-integral Method: Allowing Nearlying Transition Points},
(Springer, New York, 1996).

\bibitem{Froman2}
N.~ Froman and P.~O.~Froman,
\emph{Physical Problems Solved by the Phase-Integral Method}
(Cambridge University Press,  Cambridge, 2005).

\bibitem{Peierls}
R.~Peierls, 
\emph{Surprises in Theoretical Physics},
(Princeton, Princeton, 1979). See esp. pp. 21-22.




\bibitem{Langer}
R.~E.~Langer,
\emph{On the connection formulas and the solutions of the wave equation},
Physical Review, {\bf51} (1937) 669--676.





\bibitem{Berry0}
M.~V.~Berry,
\emph{Uniform approximation: A new concept in wave theory},
Sci. Prog., Oxf.   {\bf 57} (1969), 43--64.

\bibitem{Berry1}
M.~V.~Berry and K.~E.~Mount,
\emph{Semiclassical approximations in wave mechanics},
Reports of progress in physics {\bf35} (1972) 315--397.

\bibitem{Arnold}
J.~M.~Arnold,
\emph{Inhomogeneous dielectric waveguides: a uniform asymptotic theory},
J. Phys. A: Math. Gen. {\bf 13} (1980) 347--360.

\bibitem{Khorasani}
S.~Korasani and A.~Adibi,
\emph{Analytical solution of linear ordinary differential equations by a differential transfer matrix method},
Electronic Journal of Differential Equations {\bf79} (2003) 1--18.






\bibitem{Bounds0}
M.~Visser,
  ``Some general bounds for 1-D scattering'',
  Phys.\ Rev.\  A {\bf59}  (1999) 427--438
  [arXiv: quant-ph/9901030].
  
\bibitem{Bounds-beta}
P. Boonserm and M. Visser,
``Bounding the Bogoliubov coefficients'',
Annals of Physics {\bf323} (2008) 27792798
[arXiv: quant-ph/0801.0610].
  
\bibitem{Bounds-greybody}
  P.~Boonserm and M.~Visser,
  ``Bounding the greybody factors for Schwarzschild black holes,''
  Phys.\ Rev.\  D {\bf 78} (2008) 101502
  [arXiv:0806.2209 [gr-qc]].
  
 \bibitem{Bounds-mg}
  P.~Boonserm and M.~Visser,
  ``Transmission probabilities and the Miller-Good transformation,''
  J.\ Phys.\ A  {\bf 42} (2009) 045301
  [arXiv:0808.2516 [math-ph]].

 
\bibitem{Bounds-analytic}
  P.~Boonserm and M.~Visser,
  ``Analytic bounds on transmission probabilities,''
  arXiv:0901.0944 [math-ph].
  
 \bibitem{Bounds-thesis}
 P.~Boonserm,
 \emph{Rigorous bounds on Transmission, Reflection, and Bogoliubov coefficients},
 (PhD thesis),   arXiv:0907.0045 [math-ph]
   


\bibitem{Bordag}
M.~Bordag,~J.~Lindig, and ~V.~M.~Mostepaneko, 
``Particle creation and vacuum polarization of a non-conformal scalar field near the isotropic cosmological singularity'',
Class.\ Quantum.\ Grav.\ {\bf 15} 581 (1998).


\bibitem{Transfer}
Transfer matrix techniques are discussed, at varying levels of detail, in the textbooks by Merzbacher~\cite{Merzbacher}, Singh~\cite{Singh}, and Mathews and Venkatesan~\cite{Mathews}, and also in the research article by Khorasani and Adibi~\cite{Khorasani}.

\end{thebibliography}
\end{document}